\documentstyle[11pt,amssymb]{article}
\begin{document}
\newcommand{\iint}{{\int\hskip-3mm\int}}
\newcommand{\dddot}[1]{{\mathop{#1}\limits^{\vbox to 0pt{\kern 0pt
 \hbox{.{\kern-0.25mm}.{\kern-0.25mm}.}\vss}}}}

\renewcommand{\theequation}{\thesection.\arabic{equation}}
\let\ssection=\section
\renewcommand{\section}{\setcounter{equation}{0}\ssection}

\newcommand{\be}{\begin{equation}}
\newcommand{\ee}{\end{equation}}
\newcommand{\bes}{\begin{eqnarray}}
\newcommand{\ees}{\end{eqnarray}}
\newcommand{\eens}{\nonumber\end{eqnarray}}

\newcommand{\qed}{{\hbox{$\  \Box$}}}

\newtheorem{theorem}{Theorem}[section]
\newtheorem{lemma}[theorem]{Lemma}
\newtheorem{proposition}[theorem]{Proposition}
\newtheorem{definition}[theorem]{Definition}

\newcommand{\dt}[1]{\dot#1}
\newcommand{\ddt}[1]{\ddot#1}
\newcommand{\dddt}[1]{\dddot#1}

\newcommand{\nl}{\nonumber\\}
\newcommand{\bl}{&&\qquad}
\newcommand{\ab}{\allowbreak}
\renewcommand{\/}{\over}
\renewcommand{\d}{\partial}
\newcommand{\dlt}{\delta}
\newcommand{\si}{\sigma}
\newcommand{\om}{\omega}
\newcommand{\la}{\lambda}
\newcommand{\ka}{\kappa}
\newcommand{\e}{{\rm e}}
\newcommand{\tr}{{\rm tr}}
\newcommand{\ext}{{\rm ext}}
\newcommand{\xcy}{{\xi\leftrightarrow\eta}}
\newcommand{\tp}{\tilde p}

\newcommand{\tdiff}{{\widetilde{diff}}}
\newcommand{\tmap}{{\widetilde{map}}}
\newcommand{\toplus}{{\widetilde{\oplus}}}
\newcommand{\tltimes}{{\widetilde{\ltimes}}}

\newcommand{\dx}{{d^Nx}}
\newcommand{\xmu}{\xi^\mu}
\newcommand{\ynu}{\eta^\nu}
\newcommand{\zsi}{\zeta^\si}
\newcommand{\xz}{\xi^0}
\newcommand{\yz}{\eta^0}
\newcommand{\zz}{\zeta^0}
\newcommand{\dmu}{{\d_\mu}}
\newcommand{\dnu}{{\d_\nu}}
\newcommand{\dsi}{{\d_\si}}
\newcommand{\dtau}{{\d_\tau}}
\newcommand{\drho}{{\d_\rho}}
\newcommand{\qmu}{{\dt q^\mu}}
\newcommand{\qnu}{{\dt q^\nu}}
\newcommand{\qsi}{{\dt q^\si}}
\newcommand{\qtau}{{\dt q^\tau}}
\newcommand{\qrho}{{\dt q^\rho}}
\newcommand{\pnu}{p_\nu}
\newcommand{\mum}{{\mu_1..\mu_m}}
\newcommand{\nun}{{\nu_1..\nu_n}}

\newcommand{\txi}{{\tilde\xi}}
\newcommand{\teta}{{\tilde\eta}}
\newcommand{\Lz}{{\cal L}^0} 
\newcommand{\Lzxi}{\Lz_\xi} 
\newcommand{\Lzeta}{\Lz_\eta}
\newcommand{\nnn}{{1\/2\pi}} 
\newcommand{\nnni}{{1\/2\pi i}} 
\newcommand{\half}{{1\/2}}

\renewcommand{\L}{{\cal L}}
\newcommand{\J}{{\cal J}}
\newcommand{\MM}{{\cal M}}
\newcommand{\FF}{{\cal F}}
\newcommand{\Lxi}{\L_\xi}
\newcommand{\Leta}{\L_\eta}
\newcommand{\oj}{{\frak g}}

\newcommand{\bxy}{{[\xi,\eta]}}
\newcommand{\ket}[1]{\big{|}#1\big{\rangle}}

\newcommand{\km}[1]{{\widehat{#1}}}
\newcommand{\no}[1]{{:\!#1\!:}}

\newcommand{\RR}{{\Bbb R}}
\newcommand{\CC}{{\Bbb C}}
\newcommand{\ZZ}{{\Bbb Z}}
\newcommand{\NN}{{\Bbb N}}

\topmargin 1.0cm

\newpage
\vspace*{-3cm}
\pagenumbering{arabic}
\begin{flushright}
{\tt physics/9705040}
\end{flushright}
\vspace{12mm}
\begin{center}
{\huge Lowest-energy representations of non-centrally extended 
 diffeomorphism algebras}\\[14mm]
\renewcommand{\baselinestretch}{1.2}
\renewcommand{\footnotesep}{10pt}
{\large T. A. Larsson\\
}
\vspace{12mm}
{\sl Vanadisv\"agen 29\\
S-113 23 Stockholm, Sweden}\\
email: tal@hdd.se
\end{center}
\vspace{3mm}
\begin{abstract}
We describe a class of non-central extensions of the
diffeomorphism algebra in $N$-dimensional spacetime, and
construct lowest-energy modules thereof, thus 
generalizing work of Eswara-Rao and Moody. There is one representation
for each representation of $Vir\tltimes_{k_0}\km{gl(N)}$ 
(an extension of the semi-direct product).
Similar modules are constructed for gauge algebras.
\end{abstract}
\vspace{1cm}
\renewcommand{\baselinestretch}{1.5}

\section{Introduction}
Let $diff(N)$ denote the diffeomorphism algebra in $N$-dimensional
spacetime. In a significant paper, Eswara-Rao and Moody constructed the 
first interesting lowest-energy representations of a non-central 
extension thereof \cite{ERM94}, but they failed to explicitly describe
the extension (except for the ``spatial'' subalgebra generated by
time-independent vector fields). In the present paper,
a four-parameter non-central extension 
$\tdiff(N; c_1,c_2,c_3,c_4)$ is described explicitly,
and a realization of this algebra is constructed
for each representation of $Vir\tltimes_{k_0}\km{gl(N)}$
(an extension of $Vir\ltimes\km{gl(N)}$).
The representations in \cite{ERM94} are recovered (in a Fourier basis)
by picking a particular vertex operator module for $Vir$ and 
the trivial module for $\km{gl(N)}$. 
Thus, my results are related to theirs 
in about the same way as tensor densitites are related to functions.
Similar results also hold for the algebra of gauge transformations 
on spacetime; special cases were previously found by 
\cite{EMY92,FM94,MEY90}. A supersymmetric generalization of the present
work can be found on the web \cite{Lar97}.

After this work was completed I became aware of related references
\cite{BB98,Bil97}.

\section{Extension}
Let $\xi=\xmu\dmu$ be a vector field and $\Lxi$ the 
Lie derivative. Greek indices $\mu,\nu = 0,1,..,N-1$ label spacetime
coordinates and the summation convention is used.
The diffeomorphism algebra (algebra of vector fields, Witt algebra) 
$diff(N)$ is generated by Lie derivatives satisfying
$[\Lxi,\Leta] = \L_\bxy$. 
Define two families of operators $S_n^\nun(g_\nun)$ and
$R_n^{\rho|\nun}(h_{\rho|\nun})$, where $g_\nun$ and $h_{\rho|\nun}$ are
arbitrary functions on spacetime. The operators are linear in the
arguments and totally symmetric in the indices $\nun$.
The following relations (\ref{Sn}--\ref{tdiff}) define a Lie algebra
extension of $diff(N)$, denoted by $\tdiff(N;c_1,c_2,c_3,c_4)$.
\bes
&&{[}\Lxi, S_n^\nun(g_\nun)] = S_n^\nun(\xmu\dmu g_\nun
 + \sum_{j=1}^n \d_{\nu_j}\xmu g_{\nu_1..\mu..\nu_n}) \nl
\bl- (n-1) S_{n+1}^{\mu\nun}(\dmu\xz g_\nun), \nl
&&S_1^\nu(\dnu f) \equiv 0, 
\label{Sn} \\
&&S^{0\nun}_{n+1}(g_\nun) =S^{\nun}_n(g_\nun), \nl
&&S_0(f) ={1\/2\pi i} \int dt f(t), \qquad
\hbox{ if $f(t)$ depends on time only,} \nl
\nl
&&{[}\Lxi, R_n^{\rho|\nun}(h_{\rho|\nun})] = 
R_n^{\rho|\nun}(\xmu\dmu h_{\rho|\nun} + \drho\xmu h_{\mu|\nun} \nl
\bl + \sum_{j=1}^n \d_{\nu_j}\xmu h_{\rho|\nu_1..\mu..\nu_n}) \nl
\bl- (n+1) R_{n+1}^{\rho|\mu\nun}(\dmu\xz h_{\rho|\nun})
 -R_{n+1}^{\rho|\mu\nun}(\drho\xz h_{\mu|\nun}) \nl
\bl+ S_{n+2}^{\rho\si\nun}(\drho\dsi\xmu h_{\mu|\nun})
 - S_{n+3}^{\rho\si\mu\nun}(\drho\dsi\xz h_{\mu|\nun}),
\label{Rn} \\
&&S_{n+1}^{\mu\nun}(\dmu h_\nun)  +
 \sum_{j=1}^n 
R_{n-1}^{\nu_j|\nu_1..\check\nu_j..\nu_n} (h_\nun) \equiv 0, \nl
\bl\hbox{where $h_\nun$ is symmetric in $\nun$.} \nl
&&R_{n+1}^{\rho|0\nun}(h_{\rho|\nun}) \equiv  
 R_n^{\rho|\nun}(h_{\rho|\nun}), \nl
&&R_n^{0|\nun}(h_\nun) \equiv 0, \nl
\nl
&&{[}\Lxi, \Leta] = \L_\bxy 
 + c_1 S_1^\rho(\drho\dnu\xmu\dmu\ynu) 
 + c_2 S_1^\rho(\drho\dmu\xmu\dnu\ynu) \nl
\bl + c_3( R_1^{\mu|\nu}(\dmu\xz\dnu\yz) + 
 S_3^{\mu\nu\rho}(\drho\dmu\xz \dnu\yz) ) \nl
\bl+ {c_4\/2} S_2^{\rho\si}( \drho\yz\dsi\dmu\xmu 
- \drho\xz\dsi\dnu\ynu )
\label{tdiff} \\
\bl+ a_1 ( S_2^{\rho\si}( \drho\dsi\xmu\dmu\yz
 -\drho\dsi\ynu\dnu\xz ) \nl
\bl- S_3^{\rho\mu\nu}(\drho\dmu\xz\dnu\yz - \drho\dnu\yz\dmu\xz) ) \nl
\bl- a_2 S_1^\rho(\drho\xz\yz) 
+ a_3  S_1^\rho(\drho\yz\dmu\xmu-\drho\xz\dnu\ynu), \nl
&&{[}S_m^\mum(g_\mum), S_n^\nun(h_\nun)] =
[S_m^\mum(g_\mum), R_n^{\rho|\nun}(h_{\rho|\nun})]  \nl
&&= [R_m^{\rho|\mum}(h_{\rho|\mum}), R_n^{\si|\nun}(h_{\si|\nun})] 
= 0. \nl
\eens
Verification of all Jacobi identities is straightforward; that these
equations define a Lie algebra also follows from the explicit realization
in theorem \ref{Lthm} below. Indeed, the extensions were discovered by
working out which algebra is generated by (\ref{Lxi}).
The extensions $a_1$ -- $a_3$ (cocycles are labelled by the factors
multiplying them) are cohomologically trivial
and may be eliminated by the redefinition
\be
\Lxi' = \Lxi + a_1 S_2^{\mu\nu}(\dmu\dnu\xz) + {a_2\/2} S_0(\xz)
 + a_3S_0(\dmu\xmu).
\ee
The remaining extensions are non-trivial, which is most easily seen
by restricting to the ``temporal'' subalgebra generated by 
vector fields on the line $x^\mu=t\dlt^\mu_0$; it is a Virasoro algebra
with central charge $12(c_1+c_2+c_3+c_4)$. In particular, this is
the whole story in one dimension, and hence (\ref{Sn}--\ref{tdiff}) 
is a natural higher-dimensional generalization of the Virasoro algebra.
The extensions $c_1$ and $c_2$ were first described by
Eswara-Rao and Moody \cite{ERM94} and myself \cite{Lar89,Lar91}, 
respectively, while $c_3$ and $c_4$ are new.  
$S_1^\rho(g_\rho)$ is a linear operator acting on one-forms 
$g_\rho dx^\rho$, 
and as such it may be viewed as a closed one-chain on spacetime. 
One possibility is that it is an exact one-chain:
\bes
S_1^\rho(g_\rho) &=& C^{\nu\rho}(\dnu g_\rho), \nl
C^{\nu\mu}(j_{\mu\nu}) &=& -C^{\mu\nu}(j_{\mu\nu}) 
\label{exact} \\
{[}\Lxi, C^{\nu\rho}(j_{\nu\rho})] &=& 
 C^{\nu\rho}(\xmu\dmu j_{\nu\rho} + \dnu\xmu j_{\mu\rho} 
  + \drho\xmu j_{\nu\mu}).
\eens
Dzhumadil'daev \cite{Dzhu96} has given an list of $diff(N)$
extensions by irreducible modules, but it seems that the extension $c_1$ 
is missing; however, it is essentially $\psi^W_1-\psi^W_3+\psi^W_4$ 
in his notation. Moreover, $c_2$ is $\psi^W_1$ and
$c_1$ and $c_2$ become $\psi^W_4$ and $\psi^W_3$ upon the substitution 
(\ref{exact}), respectively.
The remaining two cocycles are not covered by his theorem, however,
because they are extensions by reducible but indecomposable modules.

\section{Realization}
Consider the Heisenberg algebra generated by operators
$q^i(s)$, $p_j(t)$, $s, t\in S^1$, where latin indices
$i,j=1,..,N-1$ run over spatial coordinates only.
\bes
[p_j(s), q^i(t)] &=& \dlt^i_j \dlt(s-t), \nl
{[}p_i(s),p_j(t)] &=& [q^i(s), q^j(t)] = 0.
\ees
These operators can be expanded in a Fourier series; e.g.,
\be
p_j(t) = \sum_{n=-\infty}^\infty\hat p_j(n) \e^{int}.
\label{pj}
\ee 
This algebra has a Fock module $\FF$ ($\ZZ$-graded by the 
frequency $n$) generated by finite strings in the non-negative 
Fourier modes of $q^i(t)$ and the positive modes of $p_j(t)$. 
Define time components by $q^0(t) = t$ and $p_0(t) = -\dt q^i(t)p_i(t)$;
in an obvious notation,  $q^\mu(t) = (t,q^i(t))$, etc.
The following relations hold.
\bes
[q^\mu(s), q^\nu(t)] &=& 0, \nl
{[}\pnu(s), q^\mu(t)] 
 &=& (\dlt^\mu_\nu - \qmu(s)\dlt^0_\nu)\dlt(s-t), 
\label{DB}\\
{[}p_\mu(s), \pnu(t)] &=& (\dlt^0_\mu \pnu(s) 
 + \dlt^0_\nu p_\mu(t)) \dt\dlt(s-t).
\eens
Normal ordering is necessary to remove infinites and to obtain a well 
defined action of diffeomorphisms on $\FF$. 
For any function of $q(t)$ and its derivatives, let 
\be
\no{f(q(t),\dt q(t))p_j(t)} \equiv 
  f(q(t),\dt q(t))p_j^\leq(t) + p_j^>(t)f(q(t),\dt q(t)),
\ee
where $p_j^>(t)$ ($p_j^\leq(t)$) is the sum (\ref{pj}) over positive 
(non-positive) Fourier modes only.
Let $L(s)$ and $T^\mu_\nu(t)$ generate the following algebra 
$Vir_c\ab\tltimes_{k_0}\ab\km{gl(N)}_{k_1,k_2}$.
\bes
{[}L(s), L(t)] &=& (L(s) + L(t)) \dt\dlt(s-t)
+ {c\/ 24\pi i} (\dddt\dlt(s-t) + \dt\dlt(s-t)), \nl
{[}L(s), T^\mu_\nu(t)] &=& T^\mu_\nu(s) \dt\dlt(s-t)
+ {k_0\/4\pi i} \dlt^\mu_\nu\ddt\dlt(s-t), 
\label{Virgl}\\
{[}T^\mu_\nu(s), T^\si_\tau(t)] &=& ( \dlt^\si_\nu T^\mu_\tau(s) 
 - \dlt^\mu_\tau T^\si_\nu(s) )\dlt(s-t) \nl
&&- {1\/2\pi i} ( k_1 \dlt^\mu_\tau \dlt^\si_\nu
 + k_2\dlt^\mu_\nu \dlt^\si_\tau ) \dt\dlt(s-t) 
\eens
\begin{theorem}\label{Lthm}
Under the conditions above, the following expressions
\bes
&&\Lxi = \int dt\ \no{\xmu(q(t)) p_\mu(t)} 
  + \xz(q(t)) L(t) + \dnu\xmu(q(t)) T^\nu_\mu(t) \nl
&&\quad\equiv  \int dt\ \no{\xi^i(q(t)) p_i(t)} 
 - \no{\xz(q(t))\dt q^i(t) p_i(t)} \nl
\bl+ \xz(q(t)) L(t) + \dnu\xmu(q(t)) T^\nu_\mu(t),
\label{Lxi} \\
&&S_n^\nun(g_\nun) ={1\/2\pi i} \int dt\ 
 \dt q^{\nu_1}(t) ... \dt q^{\nu_n}(t) g_\nun(q(t)), \nl
&&R_n^{\rho|\nun}(h_{\rho|\nun}) = {1\/2\pi i} \int dt\ \ddt q^\rho(t) 
\dt q^{\nu_1}(t) ... \dt q^{\nu_n}(t) h_{\rho|\nun}(q(t)), 
\eens
realize the Lie algebra $\tdiff(N; 1+k_1,k_2,-2+(c+2N-2)/12,1+k_0)$, 
while the cohomologically trivial parameters are 
$a_1=-1$, $a_2=(c+2N-2)/12$, $a_3=i/2$.
\end{theorem}
The proof is deferred to the appendix.
Consequently, this algebra acts on $\FF\otimes\MM$ for every 
$Vir_c\tltimes_{k_0}\km{gl(N)}_{k_1,k_2}$ module $\MM$.
It should be stressed that this action
is manifestly well defined, at least for the subalgebra of 
vector fields that are polynomial in the spatial coordinates and a Fourier
polynomial in $x^0$, because finiteness is preserved when all operators
in (\ref{Lxi}) act on finite strings in non-negative Fourier modes
in that case.
The Hamiltonian
\be
\L_{-i\d_0} = -i\int dt\ ( -\no{ \dt q^i(t) p_i(t) } + L(t) ) 
\ee
is the operator responsible for computing the $\ZZ$-grading.

In the absense of normal ordering and central charges in (\ref{Virgl}),
(\ref{Lxi}) yields a proper realization of $diff(N)$. 
The higher-dimen\-sional analogue of a primary field depends on five
parameters $\la$, $w$ (defined up to an integer), 
$\ka$, $p$, and $q$:
\bes
&&[\Lxi,\phi^{\si_1..\si_p}_{\tau_1..\tau_q}(t)] =
- \xz(q(t)) \dt\phi^{\si_1..\si_p}_{\tau_1..\tau_q}(t) 
 -\la\dt\xz(q(t)) \phi^{\si_1..\si_p}_{\tau_1..\tau_q}(t) \nl
\bl+ iw \xz(q(t)) \phi^{\si_1..\si_p}_{\tau_1..\tau_q}(t)
-\ka\dmu\xmu(q(t)) \phi^{\si_1..\si_p}_{\tau_1..\tau_q}(t) \\
\bl+ \sum_{i=1}^p \dmu \xi^{\si_i}(q(t)) 
 \phi^{\si_1..\mu..\si_p}_{\tau_1..\tau_q}(t)
- \sum_{j=1}^q \d_{\tau_j} \xmu(q(t)) 
 \phi^{\si_1..\si_p}_{\tau_1..\mu..\tau_q}(t),
\eens
where $[\Lxi, q^\mu(t)] = \xmu(q(t)) - \qmu(t)\xz(q(t))$.

The result of Eswara-Rao and Moody \cite{ERM94} is recovered 
as follows: they work in a  Fourier basis on the torus, and denote
$q^i(t) = \dlt_i(z)$ and $p_j(t) = d_j(z)$, where $z = \exp(it)$.
A standard vertex operator realization for the Virasoro generator
$L(t)$ was given, based on the remaining roots $\alpha_p$, but they
missed the appearance of $\km{gl(N)}$.
Consequently, $T^\mu_\nu(t)=0$ and $k_0 = k_1 = k_2 = 0$ in their
work.

\section{Gauge algebras}
Consider the gauge algebra $map(N,\oj)$, i.e. maps from
$N$-dimensional spacetime to a finite-dimensional Lie algebra $\oj$,
where $\oj$ has basis $J^a$, structure constants $f^{ab}{}_c$, and
Killing metric $\dlt^{ab}$. Define constants $g^a$ and ${g'}^a$ 
satisfying $f^{ab}{}_c g^c = f^{ab}{}_c {g'}^c = 0$. Clearly, 
$g^a = {g'}^a = 0$ if $J^a\in[\oj,\oj]$, but they may be non-zero on
abelian factors. Let $X=X_a(x)J^a$, $x\in\RR^N$ be a $\oj$-valued
function and define $[X,Y]_c = if^{ab}{}_c X_aY_b$.
$diff(N)\ltimes map(N,\oj)$ has the non-central extension
$\tdiff(N;c_1,c_2,c_3,c_4)\ab\tltimes_{g,g'}\ab
\tmap(N,\oj;k)$, with brackets
\bes
[\J_X, \J_Y] &=& \J_{[X,Y]} - k \dlt^{ab} S_1^\rho(\drho X_aY_b), \nl
{[}\Lxi, \J_X] &=& \J_{\xmu\dmu X} 
 - g^a S_2^{\mu\nu}(\dmu\xz\dnu X_a) 
- {g'}^a S_1^\rho(\drho\dmu\xmu X_a), \nl
{[}\J_X, S_n^\nun(g_\nun)] &=& 
{[}\J_X, R_n^{\rho|\nun}(h_{\rho|\nun})] = 0,
\label{tmap}
\ees
in addition to (\ref{tdiff}).
Let $J^a(t)$, $t\in S^1$, generate the Kac-Moody algebra $\km{\oj}_k$.
Consider the algebra 
$Vir_c\tltimes_{k_0,g}(\km{gl(N)}_{k_1,k_2}\toplus_{g'}\km{\oj}_k)$,
with brackets (\ref{Virgl}) and
\bes
[J^a(s),J^b(t)] &=& if^{ab}{}_c J^c(s) \dlt(s-t) 
 + {k\/2\pi i} \dlt^{ab}\dt\dlt(s-t),  \nl
{[}T^\mu_\nu(s),J^a(t)] &=& {{g'}^a\/2\pi i} \dlt^\mu_\nu \dt\dlt(s-t),
\label{kmg} \\
{[}L(s),J^a(t)] &=& J^a(s) \dt\dlt(s-t) + {g^a\/2\pi i} \ddt\dlt(s-t).
\eens
Then
\be
\J_X = \int dt\ X_a(q(t)) J^a(t)
\ee
yields a realization of $\tmap(N,\oj;k)$, with the intertwining action of
$\tdiff(N;\ab c_1,c_2,c_3,c_4)$ described above, 
and the parameters 
$k$, $g^a$ and ${g'}^a$ in (\ref{tmap}) and (\ref{kmg}) agree.

\appendix
\section{Proof of theorem \ref{Lthm}}
We first prove that in absense of normal ordering, (\ref{Lxi}) defines
a proper realization of $diff(N)$.
The operators $\tp_\nu(t) = p_\nu(t) + \dlt^0_\nu L(t)$ satisfy
relations (\ref{DB}) and also
\be
{[}\tp_\mu(s), T^\nu_\si(t)]
= \dlt^0_\mu T^\nu_\si(s) \dt\dlt(s-t).
\ee
Introduce the abbreviated notation $\xmu(t) \equiv \xmu(q(t))$.
Now,
\bes
[\Lxi,\Leta] &=& \iint ds dt\
 [\xmu(s) \tp_\mu(s) + \dsi \xmu(s) T^\si_\mu(s),
  \ynu(t) \tp_\nu(t) + \d_\tau \ynu(t) T^\tau_\nu(t)] \nl
&=& \iint ds dt \ \xmu(s) \Big\{
\d_\rho \ynu(t) (\dlt^\rho_\mu - \dlt^0_\mu\qrho(s))
\dlt(s-t) \tp_\nu(t) \nl
&&+ \ynu(t) \dlt^0_\mu \tp_\nu(s) \dt\dlt(s-t) \Big\} \\
&& + \xmu(s) \Big\{ \d_\rho \d_\tau\ynu(t)
(\dlt^\rho_\mu - \dlt^0_\mu\qrho(s)) \dlt(s-t) T^\tau_\nu(t) \nl
&& + \d_\tau\ynu(t) \dlt^0_\mu T^\tau_\nu(s) \dt\dlt(s-t) \Big\} \nl
&& + \dsi \xmu(s) \d_\tau \ynu(t) \dlt^\tau_\mu T^\si_\nu(s)
\dlt(s-t) - \xcy,
\eens
where $\xcy$ stands for the same expression with $\xi$ and
$\eta$ interchanged everywhere. Rewrite the terms proportial
to the derivative of the delta function by noting that
\be
\iint ds dt \ f(s)g(t) \dt\dlt(s-t)
= \int f \dt g = -\int \dt f g.
\ee
The function arguments were suppressed in the single integrals, because no
confusion is possible.
This leaves us with
\bes
&& \int \xmu (\dmu \ynu - \dlt^0_\mu \qrho\d_\rho\ynu) \tp_\nu
 + \xmu \dt\ynu \dlt^0_\mu \tp_\nu \nl
&& + \xmu\d_\rho \d_\tau\ynu
  (\dlt^\rho_\mu - \dlt^0_\mu\qrho)T^\tau_\nu
 + \xmu \d_\tau\dt\ynu \dlt^0_\mu T^\tau_\nu
 + \dsi \xmu \dmu \ynu T^\si_\nu - \xcy \nl
&=& \int \xmu \dmu \ynu \tp_\nu + \xmu \dmu \d_\tau\ynu T^\tau_\nu
 + \dsi \xmu \dmu \ynu T^\si_\nu - \xcy \\
&=& \int (\xmu \dmu \eta)^\nu \tp_\nu
 + \d_\tau(\xmu \dmu \ynu) T^\tau_\nu - \xcy
= \L_{\xmu \dmu \eta}- \xcy,
\eens
where we used that $\dt\ynu = \qrho\d_\rho\ynu$.
Hence $[\Lxi, \Leta] = \L_\bxy$, and it is clear that normal ordering
must result in some abelian extension of $diff(N)$.
We now proceed to calculate it.

Split the delta function into positive and negative energy parts.
\be
\dlt^>(t) = \nnn \sum_{m>0} \e^{-imt}, \qquad
\dlt^\leq (t) = \nnn \sum_{m\leq0} \e^{-imt}.
\ee
\begin{lemma}{\label{deltalemma}} 
\bes
&i.& \dlt^>(t) \dlt^\leq (-t) - \dlt^>(-t) \dlt^\leq (t)
= -\nnni\dt \dlt(t) \nl
&ii.& \dlt^>(t) \dt\dlt^\leq (-t) - \dt\dlt^>(-t) \dlt^\leq (t)
= {1\/ 4\pi i} (\ddt \dlt(t) + i\dt \dlt(t)) \nl
&iii.& \dt\dlt^>(t) \dt\dlt^\leq (-t) - \dt\dlt^>(-t) \dt\dlt^\leq (t)
= {1\/ 12\pi i} (\dddt \dlt(t) + \dt \dlt(t)) 
\eens
\end{lemma}
{\em Proof:}
\bes
&i.&\ 4\pi^2 \cdot LHS =
\sum_{m>0} \sum_{n\leq0} ( \e^{-i(m-n)t} - \e^{i(m-n)t} )
= \sum_{k>0} \sum_{m=1}^k (\e^{-ikt} - \e^{ikt}) \nl
&&= \sum_{k>0} k (\e^{-ikt} - \e^{ikt})
= \sum_k k \e^{-ikt} = 2\pi i \dt\dlt(t)
\hbox{ where } k = m-n. \nl
&ii.&\ 4\pi^2 i \cdot LHS =
\sum_{m>0} \sum_{n\leq0} ( n\e^{-i(m-n)t} - m\e^{i(m-n)t} ) \nl
&&= \sum_{k>0} \sum_{m=1}^k (m-k)\e^{-ikt} - m \e^{ikt}
= \sum_{k>0} -{k(k-1)\/ 2} \e^{-ikt} - {k(k+1)\/ 2} \e^{ikt} \nl
&&= -\sum_k {k(k-1)\/ 2} \e^{-ikt}
= \pi \ddt\dlt(t) + \pi i\dt\dlt(t). \nl
&iii.&\ -4\pi^2 \cdot LHS =
\sum_{m>0} \sum_{n\leq0} ( mn\e^{-i(m-n)t} - mn\e^{i(m-n)t} ) \nl
&&= \sum_{k>0} \sum_{m=1}^k m(m-k) (\e^{-ikt} - \e^{ikt})
= \sum_{k>0} -{k^3-k\/ 6} (\e^{-ikt} - \e^{ikt} )\nl
&&= -\sum_k {k^3-k\/ 6} \e^{-ikt}
= {1\/6} (2\pi i \dddt\dlt(t) + 2\pi i\dt\dlt(t)).
\nonumber
\qed
\ees
Define
\bes
\txi^i(t) &\equiv&\txi^i(q(t),\dt q(t)) 
 = \xi^i(q(t)) -\xz(q(t)) \dt q^i(t), 
\label{txi} \\
\chi^{>i}_{\xi j}(t,s) 
&\equiv& [p^>_j(t), \txi^i(s)] \nl
&=& \d_j \txi^i(s) \dlt^>(t-s) + \dlt^i_j \xz(s) \dt \dlt^>(t-s),
\ees
and $\chi^{\leq i}_{\xi j}(t,s)$ analogously.
Moreover, set  $\chi^{\ i}_{\xi j}(t,s) 
= \chi^{>i}_{\xi j}(t,s) + \chi^{\leq i}_{\xi j}(t,s)$.
\begin{lemma} \label{txilemma}
The expressions defined in (\ref{txi}) satisfy the following relations. 
\bes
\d_i\txi^i  &=& \dmu\xmu - \dt \xz
\label{dtxi} \\
 \d_j\dt\txi{}^i \d_i\teta^j &=& \dnu\dt\xmu\dmu\ynu 
 + \dnu\xz\qrho\drho\dt\ynu  
- \qrho\drho\dt\xmu\dmu\yz - \ddt\xz\dt\yz \nl
&&- \dt\xz\qrho\drho\dt\yz +\qrho\drho\dt\xz\dt\yz
 + {d\/dt}(\dt\xz\dt\yz - \dnu\xz\dt\ynu). 
\label{dydx}
\ees
\end{lemma}
{\em Proof: }
We use that $\txi^0 \equiv 0$.
Eq. (\ref{dtxi}) thus equals
\be
\dmu\txi^\mu = \dmu\xmu - \dmu\xz\qmu, \ee
whereas (\ref{dydx}) becomes
\bes
\dnu\dt\txi{}^\mu \dmu\teta^\nu 
&=& (\dnu\dt\xmu-\dnu\xz\ddt q^\mu - \dnu\dt\xz\qmu)
(\dmu\ynu - \dmu\yz\qnu) \nl
&=& \dnu\dt\xmu\dmu\ynu 
 - \dnu\xz(\ddt\ynu-\qrho\drho\dt\ynu) - \dnu\dt\xz\dt\ynu \nl
 &&- \qrho\drho\dt\xmu\dmu\yz + \dt\xz(\ddt\yz-\qrho\drho\dt\yz) 
  + \qrho\drho\dt\xz\dt\yz.
\qed
\ees
Consider
\be
\Lzxi = \int dt\ \no{ \xmu(q(t)) p_\mu(t) } 
\equiv \int dt\ (\txi^i(t) p_i^\leq (t) + p_i^>(t) \txi^i(t)).
\ee
\bes
[\Lzxi, \Lzeta] &=& \iint ds dt\
 [\txi^i(s) p_i^\leq (s) + p_i^>(s) \txi^i(s), \nl
 \bl\teta^j(t) p_j^\leq (t) +  p_j^>(t) \teta^j(t)] \\ 
&=& \iint ds dt\Big\{
 \txi^i(s) \chi^{\leq j}_{\eta i}(s,t)p^\leq_j(t) 
-\teta^j(t) \chi^{\leq  i}_{\xi j}(t,s)p^\leq_i(s) \nl
&& + \txi^i(s) p^>_j(t) \chi^{\leq j}_{\eta i}(s,t) 
 -\chi^{> i}_{\xi j}(t,s)\teta^j(t) p^\leq_i(s) \nl
&& + \chi^{> j}_{\eta i}(s,t)p^\leq_j(t)\txi^i(s) 
-p^>_i(s) \teta^j(t) \chi^{\leq  i}_{\xi j}(t,s) \nl 
&& -p^>_i(s) \chi^{> i}_{\xi j}(t,s)\teta^j(t) 
+p^>_j(t)\chi^{> j}_{\eta i}(s,t) \txi^i(s) \Big\}.
\eens
Of these eight terms, the third can be rewritten as
\be
p^>_j(t) \txi^i(s) \chi^{\leq j}_{\eta i}(s,t)
-\chi^{> i}_{\xi j}(t,s) \chi^{\leq j}_{\eta i}(s,t) 
\ee
and the fifth as
\be
 \chi^{> j}_{\eta i}(s,t) \txi^i(s)p^\leq_j(t)
 + \chi^{> j}_{\eta i}(s,t)\chi^{\leq  i}_{\xi j}(t,s).
\ee
Hence 
\bes
&&[\Lzxi, \Lzeta] = \iint ds dt\Big\{
 \txi^i(s) \chi^{\leq j}_{\eta i}(s,t)p^\leq_j(t)
 -\teta^j(t) \chi^{\leq  i}_{\xi j}(t,s)p^\leq_i(s) \nl 
\bl+ p^>_j(t) \txi^i(s) \chi^{\leq j}_{\eta i}(s,t) 
- \teta^j(t) \chi^{> i}_{\xi j}(t,s) p^\leq_i(s) \nl
\bl+  \txi^i(s)\chi^{> j}_{\eta i}(s,t)p^\leq_j(t) 
- p^>_i(s) \teta^j(t) \chi^{\leq  i}_{\xi j}(t,s) \\
\bl-p^>_i(s) \teta^j(t) \chi^{> i}_{\xi j}(t,s)
 +  p^>_j(t) \txi^i(s) \chi^{> j}_{\eta i}(s,t)) \nl 
\bl- \chi^{> i}_{\xi j}(t,s) \chi^{\leq j}_{\eta i}(s,t) 
+ \chi^{> j}_{\eta i}(s,t)\chi^{\leq  i}_{\xi j}(t,s) \Big\}.
\eens
The regular piece is
\be
\iint ds dt\ \txi^i(s) {\chi_\eta}^j_i(s,t) p^\leq_j(t) 
+ p^>_j(t) \txi^i(s){\chi_\eta}^j_i(s,t) - \xcy.
\ee
We focus on the first term.
\bes
&&\iint ds dt\ \txi^i(s) {\chi_\eta}^j_i(s,t) p^\leq_j(t) - \xcy \nl
&=& \iint ds dt\ \txi^\mu(s) ( \dmu\teta^j(t) \dlt(s-t)
 +\yz(t) \dlt^j_\mu \dt\dlt(s-t) ) p^\leq_j(t) - \xcy \nl
&=& \int \Big\{ (\widetilde{ \xmu\dmu\eta})^j 
 - \xz(\dt\eta^j - \dt\yz\dt q^j) - \dt\txi{}^j \yz  \Big\} p^\leq_j
 - \xcy,
\ees
which equals $\Lz_\bxy$.  We again suppress the integration variable 
in single integrals, and write $\xmu(s) \equiv \xmu(q(s))$, etc.
The extension $\ext_0(\xi,\eta)$ becomes
\bes
&& \iint ds dt \Big\{
 - \chi^{>i}_{\xi j}(t,s) \chi^{\leq j}_{\eta i}(s,t)
+ \chi^{>j}_{\eta i}(s,t) \chi^{\leq i}_{\xi j}(t,s) \Big\}\nl
&=& - \iint ds dt \Big\{ (\d_j \txi^i(s) \dlt^>(t-s)
 + \dlt^i_j \xz(s) \dt \dlt^>(t-s))\times \nl
 && \times (\d_i \teta^j(t) \dlt^\leq (s-t) 
  + \dlt^j_i \yz(t) \dt \dlt^\leq (s-t) ) \Big\} - \xcy \nl
&=& - \iint ds dt \Big\{ \d_j \txi^i(s) \d_i \teta^j(t)
 \dlt^>(t-s) \dlt^\leq (s-t) \nl
 &&+ \xz(s) \d_j \teta^j(t) \dt\dlt^>(t-s) \dlt^\leq (s-t) \nl
 &&+ \d_i \txi^i(s) \yz(t) \dlt^>(t-s) \dt\dlt^\leq (s-t) \nl
 &&+ \dlt^i_i \xz(s) \yz(t)\dt\dlt^>(t-s) \dt\dlt^\leq (s-t) \Big\}
  - \xcy \label{ext0} \\
&=& \nnni \iint ds dt \Big\{ \d_j \txi^i(s) \d_i \teta^j(t) 
 \dt\dlt(t-s)\nl
&&+ \half \xz(s) \d_j \teta^j(t) (\ddt\dlt(t-s) - i\dt \dlt(t-s)) \nl
&&- \half \d_i \txi^i(s) \yz(t) (\ddt\dlt(t-s) + i\dt\dlt(t-s)) \nl
 &&- {N-1\/6} \xz(s) \yz(t) (\dddt\dlt(t-s) + \dt \dlt(t-s)) \Big\} \nl
&=& \nnni \int \Big\{ \d_j \dt\txi{}^i \d_i \teta^j 
 - \half \dt\xz \d_j \dt\teta{}^j 
 + \half \d_i \dt\txi{}^i \dt\yz \nl
&&- {N-1\/6} ( -\ddt\xz \dt\yz + \dt\xz\yz )
+ {i\/ 2} ( -\dt\xz \d_j \teta^j+ \d_i \txi^i \dt\yz ) \Big\},
\eens
where we used Lemma \ref{deltalemma} and the fact that 
$\dlt^i_i = N-1$.
Now consider the full algebra. The regular piece follows from the
following calculation.
\bes
&&\int(\xmu\dmu\ynu-\xz\dt\ynu)\pnu
+\iint ds dt\ \xmu(s)\ynu(t)\dlt^0_\mu\pnu(s)\dt\dlt(s-t) \nl
&& + \int(\xmu\dmu\yz-\xz\dt\yz)L 
+\iint ds dt\ \xz(s)\yz(t)L(s)\dt\dlt(s-t) \nl
&& +\int \Big\{ (\xmu\dmu\dsi\eta^\tau-\xz\dsi\dt\eta^\tau)T^\si_\tau
+ \dmu\xi^\nu\dsi\eta^\tau\dlt^\si_\nu T^\mu_\tau \Big\} \nl
&&+\iint ds dt\ \xz(s)\dsi\eta^\tau(t) T^\si_\tau(s)\dt\dlt(s-t) 
-\xcy \\
&=& \int \bxy^\nu\pnu + \bxy^0L + \dmu\bxy^\nu T^\mu_\nu,
\eens
and the full extension is
\bes
\ext(\xi,\eta) &=&  \ext_0(\xi,\eta)
+ \iint ds dt\ \Big\{
{c\/24\pi i} \xz(s)\yz(t) (\dddt\dlt(s-t)+ \dt\dlt(s-t)) \nl
&& + {k_0\/4\pi i} (\xz(s)\dnu\ynu(t) 
 - \dmu\xmu(s)\yz(t)) \ddt\dlt(s-t) \nl
&&- \dsi\xmu(s) \d_\tau\ynu(t) ({k_1\/2\pi i} \dlt^\si_\nu \dlt^\tau_\mu 
+ {k_2\/2\pi i} \dlt^\si_\mu \dlt^\tau_\nu )\dt\dlt(s-t) \Big\} \\
&=& \ext_0(\xi,\eta) + \nnni \int \Big\{
 {c\/12}(\ddt\xz\dt\yz - \dt\xz\yz) \nl
&&+ {k_0\/2} (-\dt\xz\dnu\dt\ynu+\dt\yz\dmu\dt\xmu)
+ k_1 \dnu\dt\xmu\dmu\ynu + k_2 \dmu\dt\xmu\dnu\ynu \Big\}.
\eens
The result now follows by means of lemma \ref{txilemma}.
\bes
&&\ext(\xi,\eta) = \nnni \int dt\ \Big\{ 
(1+k_1) \dnu\dt\xmu \dmu\ynu + k_2 \dmu\dt\xmu \dnu\ynu \nl
\bl +\dnu\xz \qrho\drho\dt\ynu-\qrho\drho\dt\xmu \dmu\yz 
- \dt\xz \qrho\drho\dt\yz  + \qrho\drho\dt\xz \dt\yz 
\label{ext} \\
 \bl  + {1+k_0\/2}( \dmu\dt\xmu \dt\yz - \dt\xz\dnu\dt\ynu )
 - (2-{c+2(N-1)\/12}) \ddt\xz \dt\yz \nl
 \bl - {c+2(N-1)\/12}\dt\xz \yz + {i\/ 2} ( \dmu \xmu \dt\yz
  - \dt\xz \dnu\ynu ) \Big\} , 
\eens
where $\dt f = \qrho\drho f$.
As a consistency check we note that the extension satisfies
 $\ext(\eta,\xi) = - \ext(\xi,\eta)$.

To calculate the remaining brackets is a straightforward task. Note 
that normal ordering is irrelevant here, because $S_n^\nun$ and
$R_n^{\rho|\nun}$ depend on $q^\mu$ only whereas $\Lxi$ depends only
linearly on $\pnu$.
\qed

{\bf Note added. }
A. Dzhumadil'daev has explained his results \cite{Dzhu96},
which I had slightly misunderstood.
The Rao-Moody cocycle $c_1$ is included in his list; it is
equivalent to his cocycle $\psi_4^W$, with coefficients in
$\Omega^1_{DeRham}/ B^1_{DeRham} \cong B^2_{DeRham} \oplus H^1_{DeRham}.$
Similarly, $c_2$ is his $\psi_3^W$. $H^1_{DeRham}$ is an 
$N$-dimensional trivial $diff(N)$ module; setting it to zero gives
the substitution (\ref{exact}).
The closedness condition $S_1^\rho(\drho f) = 0$ can be lifted for
the cocycle $c_2$ (but not for $c_1$). One then obtains $\psi_1^W$, 
first discovered in \cite{Lar89}.
Dzhumadil'daev considered extensions by modules of tensor fields, 
not necessarily irreducible. $c_3$ and $c_4$ are not included in his
list, because they are extensions by other types of modules.


\begin{thebibliography}{99}

\bibitem{BB98} Berman, S. and Y. Billig,
  {\em Irreducible representations for toroidal Lie algebras},
  preprint (1998).

\bibitem{Bil97} Billig, Y.,
  {\em Principal vertex operator representations for toroidal
  Lie algebras},
  J. Math. Phys. {\bf 7}, 3844--3864 (1998).

\bibitem{Dzhu96} Dzhumadildaev A.,
  {\em Virasoro type Lie algebras and deformations},
  Z. Phys. C {\bf 72}, 509--517 (1996).

\bibitem{EMY92} Eswara Rao, S., R.V. Moody and T. Yokonuma,
  {\em Lie algebras and Weyl groups arising from vertex operator
  representations},
  Nova J. of Algebra and Geometry {\bf 1}, 15--57 (1992).

\bibitem{ERM94} Eswara Rao, S. and R.V. Moody,
  {\em Vertex representations for $N$-toroidal Lie algebras and a
  generalization of the Virasoro algebra},
  Commun. Math. Phys. {\bf 159}, 239--264 (1994).

\bibitem{FM94} Fabbri, M. and R.V. Moody,
  {\em Irreducible representations of Virasoso-toroidal Lie algebras},
  Commun. Math. Phys. {\bf 159}, 1--13 (1994).

\bibitem{Lar89} Larsson, T.A.,
  {\em Multi-dimensional Virasoro algebra},
  Phys. Lett. {\bf A 231}, 94--96 (1989).

\bibitem{Lar91} Larsson, T.A.,
  {\em Central and non-central extensions of multi-graded Lie algebras },
  J. Phys. A. {\bf 25}, 1177--1184 (1992).

\bibitem{Lar97} Larsson, T.A.,
  {\em Fock representations of non-centrally extended super-diffeomorphism
  algebras},
  {\tt physics/9710022} (1997).

\bibitem{MEY90} Moody, R.V., S. Eswara Rao and T. Yokonoma,
  {\em Toroidal Lie algebras and vertex representations},
  Geom. Ded. {\bf 35}, 283--307 (1990).

\end{thebibliography}
\end{document}